# Feasibility of Estimating Macroscopic Fundamental Diagrams in Iran Traffic Network

Mahgam Tabatabaei[a], Reza Golshan Khavas[a], Ali Tavakoli Kashani[a]

[a] *Department of Civil Engineering, Iran University of Science and Technology, Tehran, Iran*

## Abstract

The increase in population and economic growth, coupled with accelerated urbanization and suburbanization, has exacerbated traffic congestion and environmental challenges in urban areas. To address these issues, a comprehensive traffic management program has been introduced, aimed at enhancing the regulation and control of traffic flow, thereby ensuring faster and safer travel. By leveraging data collected from various regions, tailored traffic management strategies can be implemented to meet the specific needs of different city sectors. This approach involves continuous monitoring of traffic conditions and the application of targeted interventions to mitigate congestion and improve overall traffic efficiency. A case study in Tehran exemplifies the application of this program. A designated section of the city's traffic network is being utilized to test the program's efficacy. The objective is to assess its practical effectiveness under real-world conditions and refine the program based on empirical findings. This initiative aims to provide a robust solution to Tehran's traffic challenges, contributing to improved traffic management and enhanced safety.

# 1. Introduction

Traffic management has emerged as a critical challenge in contemporary urban environments, driven by the exponential growth in traffic flow and the increasing complexity of transportation networks. Addressing these challenges requires the development of innovative methodologies that combine traditional traffic control principles with cutting-edge technologies. Recent advancements highlight the integration of data-driven approaches, leveraging big data analytics, machine learning, and real-time monitoring systems to optimize traffic flow, reduce congestion, and enhance the sustainability of urban transportation systems (1-4). A comprehensive understanding of traffic dynamics across multiple scales—ranging from macroscopic to microscopic levels—is imperative for designing effective, adaptive traffic management strategies that can meet the demands of modern cities.

Traffic models can be categorized into three levels of detail: macroscopic, mesoscopic, and microscopic. Macroscopic traffic flow models define traffic at an aggregate level, without recognizing individual components. These models describe traffic flow characteristics using parameters like density, volume, and speed. Fundamental diagrams are applicable at both the link level and the network level. The network-level fundamental diagram, representing the relationship between macroscopic traffic variables, examines the relationship between average flow and average density in an urban network and provides a reliable framework for monitoring and controlling traffic in large cities. The initial concept of the macroscopic fundamental diagram (MFD) was introduced by Godfrey (5).

The validity of this idea was later verified through field data and simulations. In 2008, Geroliminis and Daganzo conducted an experimental study in Yokohama, using GPS data and fixed detectors, including inductive loop detectors, to demonstrate spatial homogeneity among traffic variables at the corridor level. This homogeneity indicated a well-defined relationship with minimal scatter, independent of demand, in urban networks (6).

The network-level fundamental diagram describes the relationship between accumulation (density multiplied by length) and production (flow multiplied by length). It also enables the examination of the overall relationship between total distance traveled and total vehicle hours in the network(5).

Methods to derive the network fundamental diagram can be divided into analytical inference and empirical analysis. Analytical models are limited to signalized corridors with homogeneous congestion levels, while empirical analyses are used for large and complex networks(5).

For empirical estimation of the network fundamental diagram without error, assuming full information about vehicle trajectories, Edie's definition can be applied to calculate network-wide average flow (Q), density (k), and speed (V) using a two-step method. First, the total distance traveled by individual vehicles and their travel time are aggregated over specific temporal and spatial intervals. These aggregated values are then used to calculate density and flow at the network level. This approach, however, is only feasible for microscopic simulations and is impractical due to the need for complete trip data of individual vehicles. As the fundamental diagram derived from Edie's method is based on individual vehicle data, it can serve as a reference for a network fundamental diagram (7).

By extending Edie's definition and using density and flow data reported by inductive loop detectors, the network fundamental diagram can be estimated. However, errors in this estimation may arise due to factors such as detector distribution across the network, their placement on specific links, and the selection of links equipped with detectors. For instance, detectors are often installed on highly congested links, leading to a higher observed density in the network. Similarly, detectors placed near traffic signals may overestimate network density. These issues can introduce significant scatter in the fundamental traffic diagram(8).

An alternative to inductive detectors is the use of probe vehicles, which address the challenge of incomplete detector coverage. Probe vehicles utilize data collected from mobile sensors in vehicles to provide traffic information at any point in the network. Since these devices are distributed across the entire network, their data enable comprehensive traffic condition analysis(9).

Probe vehicle data must be mapped to the road network, introducing uncertainties as exact lane-level mapping is not possible and can only align with certain road segments. Probe vehicle data are used to estimate the network fundamental diagram through simulation and empirical analysis. The penetration rate of probe vehicles (the proportion of vehicles sending data) and their spatial distribution across the network are crucial for accuracy(10).

In 2018, Knoop et al. utilized aggregated probe vehicle data provided by Google to derive a network fundamental diagram for Amsterdam. This dataset included absolute speed values for each segment and flow as a fraction of maximum flow. They calculated the aggregated speed for the network using segment speed values and estimated flow for each segment based on an assumed maximum flow(11).

While many studies rely on data from inductive loop detectors or probe vehicles to derive the network fundamental diagram, some have combined these data sources. In 2016, Ambühl and Menendez used data fusion algorithms to integrate both types of data, reducing errors in the network fundamental diagram. They developed a hybrid algorithm dividing the network into subsets, one with inductive loop detectors and the other without. By combining probe vehicle data (for network speed) with loop detector data (for traffic flow), they significantly reduced estimation errors(12).

Another data source for network fundamental diagram estimation is traffic cameras, which are primarily installed to enhance safety and control traffic. Camera images provide information about traffic flow characteristics, vehicle speed, and density within the monitored area(5).

In recent years, advancements in technology have increased the number of mobile devices used by drivers, drawing greater attention to probe vehicle data for network fundamental diagram estimation. Real-time travel time data, used in this study, were extracted from online routing platforms, which calculate travel times based on user data. To estimate the traffic network diagram for Tehran, this study utilized data from inductive loop detectors and probe vehicle travel times. Inductive detector data were used to determine vehicle volumes, while probe vehicle travel times were aggregated to derive network-level vehicle speeds.

- **Inductive Loop detector data**

In this study, inductive loop detector data from Tehran Municipality's Traffic Control Company were used. These detectors are installed at signalized intersections organized by an intelligent traffic system known as SCATS. SCATS is an Intelligent Transportation System (ITS) that dynamically schedules traffic light phases to achieve optimal timing based on current traffic conditions. The study used data outputs from the inductive loop detectors, which only reported vehicle volume. However, occupancy levels were not recorded. As a result, only aggregated vehicle volume at the network level could be calculated. The data were aggregated in 15-minute intervals, but not all intersections had access to loop detector data.

- **Probe Vehicle Travel Time Data**

This study utilized data from "Neshan," a free mobile application providing online navigation and real-time traffic information. The app calculates average speed for each road segment based on vehicle GPS data and estimates travel time accordingly. Its accuracy for travel time estimation in Tehran streets has been validated in prior studies.

Neshan offers various web services; in this study, the "distance matrix" API was used to calculate travel times between origin-destination pairs. Using travel time data, segment speed was calculated.

## 2. Methodology

- **Aggregation of inductive loop detector data**

The provided loop detector data lacked occupancy levels. Therefore, the average vehicle flow was calculated using the relationship:

$$\hat{q} = \frac{\sum_i q_i(t) l_i}{\sum_i l_i} \quad (1)$$

where $q_i(t)$ is the vehicle count on link I in time interval t, $l_i$ is the length of link I, and $\hat{q}$ is the average vehicle flow.

- **Aggregation of Probe Vehicle Travel Time Data**

Travel speed data were extracted in 5-minute intervals and aggregated into 15-minute intervals to align with the loop detector data. To do this, 96 ,15-minute intervals are considered for the 24-hour day (from 00:00 to 00:15, 00:15 to 00:30, etc.). Then, the travel times are assigned to the corresponding intervals. Finally, all travel times that fall within an interval are aggregated using equation 2 to obtain the 15-minute interval.

$$V_i = \frac{l_i}{\frac{\sum_i t_i}{n}} \quad (2)$$

where i is the length of link i, $t_i$ is the speed obtained from Neshan's distance matrix in the 5-minute time interval for that link, and n is the number of speeds recorded in that time interval. $V_i$ is the vehicle speed on link i in the 15-minute time interval. The aggregated speed for the entire network can also be calculated using equation 3. This equation is used to aggregate speed data collected from links using inductive loop data.

$$\hat{V} = \frac{\sum_i V_i l_i}{\sum_i l_i} \quad (3)$$

where $V_i$ is the vehicle speed on link i in the 15-minute time interval, obtained from equation 2 and $\hat{V}$ is the average vehicle speed for the entire network in the 15-minute time period.

- **Underwood Model Calibration**

The Underwood model is one of the single-regime traffic models that shows the relationship between traffic parameters using only one equation. This exponential model was introduced in 1960 by Underwood. The model demonstrates the relationship between speed and density using equation 4:

$$V = V_{sw} \exp\left(\frac{-k}{k_{opt}}\right) \quad (4)$$

As shown, this model has two main parameters: Vsw and kopt. Vsw is the free-flow speed, which represents the speed when traffic density and flow are close to zero. kopt is the optimal density that occurs when traffic flow is at its maximum (13). After plotting the network fundamental diagram, the Underwood model is fitted to it. The calibration of the model parameters follows these steps:

- First, both sides of the Underwood model are logarithmically transformed to convert it into a linear equation.

$$V = V_{sw} \exp\left(\frac{-k}{k_{opt}}\right) \rightarrow \ln(V) = \ln(V_{sw}) + \left(\frac{-1}{k_{opt}}\right)k \quad (5)$$

- In the second step, the model coefficients are fitted like a linear equation, and the values for this model are derived using the following relations:

$$b = \frac{\sum_{i=1}^{n}(k_i - \bar{k})(\ln(V_i) - \overline{\ln(V)})}{\sum_{i=1}^{n}(k_i - \bar{k})^2} \quad (6)$$

$$a = \overline{\ln(V)} - b\bar{k} \quad (7)$$

where $k_i$ is the density at point i, $\bar{k}$ is the average density across all observed points from 1 to n, $V_i$ is the speed at point i, and $\overline{\ln(V)}$ is the average logarithmic speed across all observed points. b is the y-intercept, which equals $\frac{-1}{k_{opt}}$, and a is equal to $\ln(V_{sw})$ (14).

## 3. Results and Discussion

As mentioned in the previous section, inductive loop data is available for one lane, while vehicle travel time data is available for a link that may have multiple lanes. In this section, for data aggregation, it is assumed that the travel time data for each lane is equal to the travel time for the entire link, and the network fundamental diagram is plotted.

The plotted network fundamental diagram is based on the aggregated speed and flow data, and then using the relation k = q/v, where k is density, q is flow, and v is speed, density is also calculated. The network fundamental diagram is also plotted based on density and flow. Figures 1 to 3 show the network fundamental diagrams obtained using aggregated inductive loop data and travel time data.

After obtaining the network fundamental diagram for speed-density, the Andrews model is fitted to it. The model parameters obtained are shown in Table 1.

As seen in the flow-density diagram, in the congested state, the scatter of points increases. In the sub-saturated state, there is less scatter, which may be due to the functional classification of the road. Since in the study network, part of Tehran, the functional classification of roads is similar, less scatter is observed in the network fundamental diagram. This finding is consistent with previous studies that examine the impact of road classification on the network fundamental diagram(15).

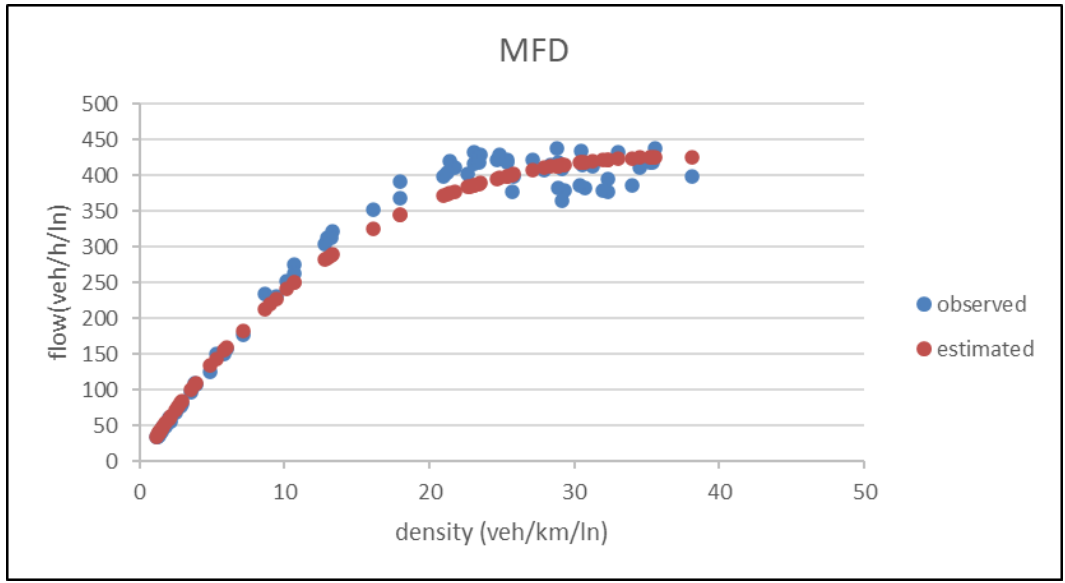

*Figure 1 MFD based on flow-density*

The network fundamental diagram based on speed and density is shown in this figure. Since in the Andrews model, speed is expressed based on density, the Andrews model is fitted to this diagram, and the results are shown in Table 1. This diagram also shows minimal scatter of points. Also, density in this diagram is derived based on the formula q = kv, where q is flow, k is density, and v is speed, with flow and speed derived from observations.

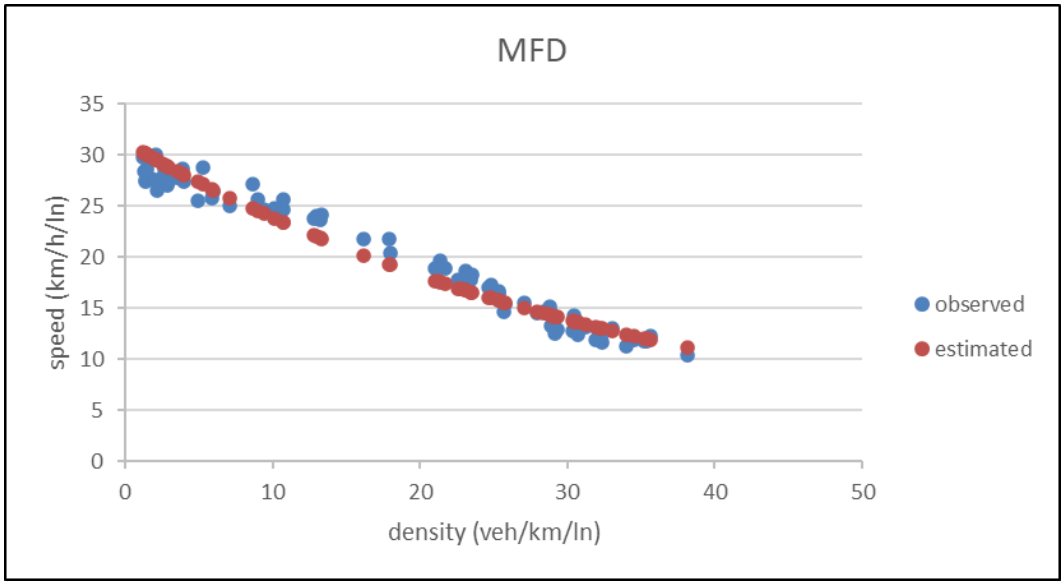

*Figure 2 MFD based on speed and density*

This diagram shows the relationship between speed and flow, both derived from observations. In this diagram, speed is based on data extracted from Neshan, and flow is based on data from the Traffic Control Company's inductive loops. As seen, scatter is observed in the saturation section of this diagram, and points in other sections also deviate further from the fitted Underwood model.

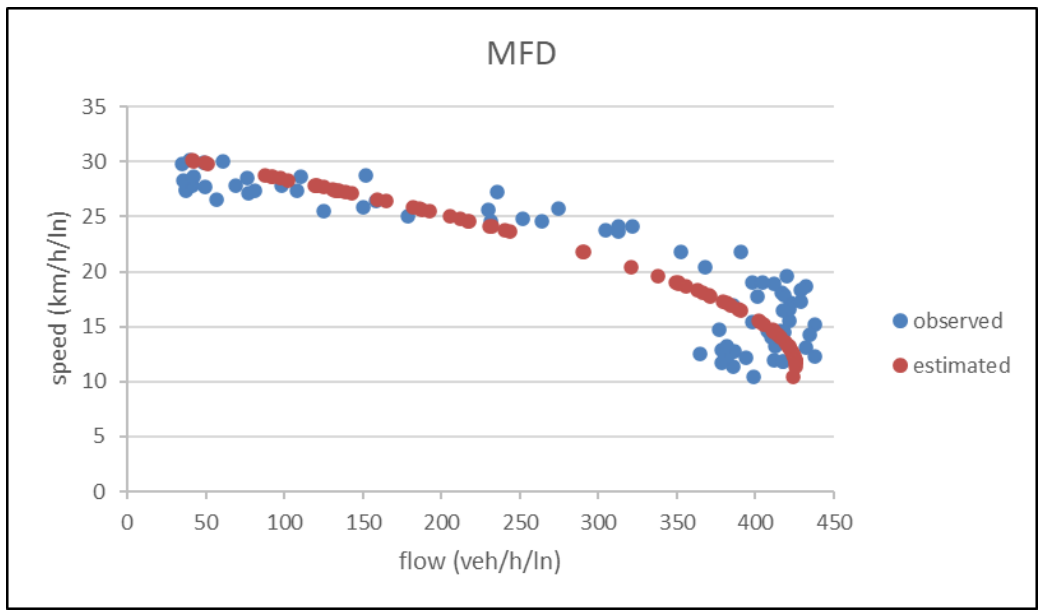

*Figure 3 MFD based on flow and speed*

Table 1 Underwood model fitted on the MFD

| Underwood Model | |
|---|---|
| Fitted Equation | $V = 31.247e^{-0.027k}$ |
| R-Squared (R²) | 0.9596 |
| Free Flow Speed (Vsw) | 31.247 km/h/ln |
| Optimal Density (Kopt) | 27.037 veh/km/ln |

## 4. Conclusion

Researchers have found that there is a relationship between three macroscopic traffic parameters on the road, namely speed, flow, and density, and by extending this relationship to a network, a tool known as the network fundamental diagram has been developed. This tool can accurately represent the status of the network and its variations. It allows one to understand the current traffic situation in the network (saturated, over-saturated, under-saturated) and predict what happens to the network when changes are made.

In this study, an attempt was made to create a network-level fundamental diagram using inductive loop data and travel time data from Neshan, and it was shown that both real-time travel time data and inductive loop data can be used to estimate the network fundamental diagram.

Given that traffic camera data is available in Tehran, future studies may incorporate this data, combining it with inductive loop data and probe vehicle travel time data, to improve the accuracy of the estimated traffic fundamental diagram. Additionally, since inductive loop data and vehicle travel time data were only analyzed over a 24-hour period in this study, it is recommended that future research extend the analysis period to a longer duration.

## 5. References

1. Kondyli A, Schrock SD, Tabatabaei M, Kummetha VC. Evaluation of Ramp Metering Effectiveness Along the I-35 Corridor in the Kansas City Metropolitan Area. Kansas Department of Transportation. Bureau of Research; 2024.

2. Afsari M, Dastmard M, Miristice LMB, Gentile G, editors. Utilizing P-median and Machine Learning for Bike-Sharing Demand Prediction. 2024 IEEE International Conference on Environment and Electrical Engineering and 2024 IEEE Industrial and Commercial Power Systems Europe (EEEIC/I&CPS Europe); 2024: IEEE.

3. Afsari M, Ippolito N, Mistrice LMB, Gentile G. Environmental benefits of taxi ride-sharing in New York City. Transportation Research Procedia. 2024;78:345-52.

4. Eldafrawi M, Varghese KK, Afsari M, Babapourdijojin M, Gentile G, editors. Predictive analytics for road traffic accidents: exploring severity through conformal prediction. Proceedings of TRB Annual Meeting 2024; 2024.

5. Zhang L, Yuan Z, Yang L, Liu Z. Recent developments in traffic flow modelling using macroscopic fundamental diagram. Transport Reviews. 2020;40(6):689-710.

6. Geroliminis N, Daganzo CF. Existence of urban-scale macroscopic fundamental diagrams: Some experimental findings. Transportation Research Part B: Methodological. 2008;42(9):759-70.

7. Leclercq L, Chiabaut N, Trinquier B. Macroscopic fundamental diagrams: A cross-comparison of estimation methods. Transportation Research Part B: Methodological. 2014;62:1-12.

8. Johari M, Keyvan-Ekbatani M, Leclercq L, Ngoduy D, Mahmassani HS. Macroscopic network-level traffic models: Bridging fifty years of development toward the next era. Transportation Research Part C: Emerging Technologies. 2021;131:103334.

9. Saffari E, Yildirimoglu M, Hickman M. A methodology for identifying critical links and estimating macroscopic fundamental diagram in large-scale urban networks. Transportation Research Part C: Emerging Technologies. 2020;119:102743.

10. Ambühl L, Loder A, Menendez M, Axhausen KW, editors. Empirical macroscopic fundamental diagrams: New insights from loop detector and floating car data. TRB 96th Annual Meeting Compendium of Papers; 2017: Transportation Research Board.

11. Knoop VL, van Erp PB, Leclercq L, Hoogendoorn SP, editors. Empirical MFDs using Google traffic data. 2018 21st International Conference on Intelligent Transportation Systems (ITSC); 2018: IEEE.

12. Ambühl L, Menendez M. Data fusion algorithm for macroscopic fundamental diagram estimation. Transportation Research Part C: Emerging Technologies. 2016;71:184-97.

13. Traffic YUBoH. Report-Yale University. Bureau of Highway Traffic1960.

14. Heckert NA, Filliben JJ, Croarkin CM, Hembree B, Guthrie WF, Tobias P, et al. Handbook 151: Nist/sematech e-handbook of statistical methods. 2002.

15. Buisson C, Ladier C. Exploring the impact of homogeneity of traffic measurements on the existence of macroscopic fundamental diagrams. Transportation Research Record. 2009;2124(1):127-36.